\documentclass[preprint,12pt,sort&compress]{elsarticle}

\usepackage[utf8]{inputenc}
\usepackage{amssymb}
\usepackage{bm}
\usepackage{color}
\usepackage{url}
\usepackage{hyperref}
\usepackage{multirow}
\usepackage{siunitx}
\newcommand{\exciting}{\texttt{exciting}}

\newcommand{\fluorine}{\texttt{fluorine}}
\newcommand{\newrt}{\texttt{fast-rt-tddft}}

\newcommand{\ie}{{\it i.e. }}
\newcommand{\eg}{{\it e.g. }}
\newcommand{\kpt}{\textbf{k}-point}

\newcommand{\kgrid}{\textbf{k}-grid}
\newcommand{\rgkmax}{\texttt{rgkmax}}

\journal{Computer Physics Communications}
\begin{document}

\title{Speeding up all-electron real-time TDDFT demonstrated by the \exciting\ package}
\author{Ronaldo Rodrigues Pela}

\address{Supercomputing Department, Zuse Institute Berlin (ZIB), Takustraße 7, 14195 Berlin, Germany}
\ead{ronaldo.rodrigues@zib.de}
\author{Claudia Draxl}
\address{Institut f\"{u}r Physik and CSMB, Humboldt-Universit\"{a}t zu Berlin, Zum Gro\ss en Windkanal 2, 12489 Berlin, Germany}
\address{European Theoretical Spectroscopic Facility (ETSF)}	
% \affiliation[1]{organization={Supercomputing Department, Zuse Institute Berlin (ZIB)},
% addressline={Takustraße 7},
% postcode={14195},
% city={Berlin},
% country={Germany}}
% \affiliation[2]{organization={Institut für Physik and CSMB, Humboldt-Universität zu Berlin},
% addressline={Zum Großen Windkanal 2},
% postcode={12489},
% city={Berlin},
% country={Germany}}
% \affiliation[2]{organization={European Theoretical Spectroscopic Facility (ETSF)},
% addressline={},
% postcode={},
% city={},
% country={}}
\date{\today}

\begin{abstract}
Currently, many \emph{ab initio} codes are being prepared for exascale computing. A first and important step is to significantly improve the efficiency of existing implementations by devising better algorithms that can accomplish the same tasks with enhanced scalability. This manuscript addresses this challenge for real-time time-dependent density functional theory in the full-potential all-electron code \exciting, with a focus on systems with reduced dimensionality. Following the strategy described here, calculations can run orders of magnitude faster than before. We demonstrate this with the molecules H$_2$ and CO, achieving speedups between 98 to over 50,000. We also present an example where conventional calculations would be particularly costly, namely the inorganic/organic heterostructure of pyridine physisorbed on monolayer MoS$_2$.
\end{abstract}
	
\maketitle
	
\section{Introduction}
In recent years, there have been increasing efforts to bring \emph{ab initio} codes to the exascale level, increasing the ultimate size of quantum systems that can be studied \cite{Gavini_2013}. This will allow the computation of more complex materials, with numbers of atoms unimaginable a few years ago. Initiatives towards new levels of scalability of \emph{ab initio} codes can be found in large-scale projects, such as the NOMAD Center of Excellence (CoE), the MAX CoE, or the TREX CoE. 

Before reaching the exascale, however, an important first step is to significantly improve the performance of a code at a single-core level, bringing the efficiency of currently implemented algorithms to a new baseline. This article addresses this challenge for real-time time-dependent density-functional theory (RT-TDDFT) as implemented in the all-electron full-potential code \exciting{} \cite{Gulans_2014,Pela_2021,Pela_2022}. Employing the (linearized) augmented planewaves plus local orbital (in short LAPW+LO) method, regarded as the gold standard for solving the Kohn-Sham (KS) equation \cite{Gulans_2018}, \exciting{} has been shown to reach ultimate precision in density-functional theory (DFT) calculations \cite{Gulans_2018} and the $GW$ approach of many-body perturbation theory \cite{Nabok_2016}. Moreover, the code is notably user-friendly, with more than 50 tutorials available on its website \cite{exciting_webpage}.

The RT-TDDFT developments presented here are mainly targeted at low-dimensional systems such as molecules, heterostructures, nanowires, and 2D materials. In these cases, codes with periodic boundary conditions, like \exciting{}, require large unit cell dimensions to avoid spurious iterations between replicas. Increasing the cell size means that more basis functions must be included in the calculations. Unfortunately, the storage of Hamiltonian and overlap matrices scales quadratically with the basis-set size, while the number of floating-point operations (FLOPs) required for matrix-matrix multiplications exhibits even cubic scaling, which obviously makes calculations heavier and slower for large unit cells. Last but not least, for evolving KS wavefunctions, large time steps cannot be afforded, since larger basis sets give rise to more KS states with higher energies. To address these challenges, we propose a twofold strategy: (i) working with submatrices to avoid explicit storage of the Hamiltonian and overlap matrices, and (ii) introducing an auxiliary subspace for evolving KS wavefunctions. By following this strategy, RT-TDDFT calculations can run orders of magnitude faster than using recent releases of \exciting{}, additionally allowing for larger time steps.

\section{Theoretical and implementation aspects}\label{sec-theory}
%\subsection{Time evolution of wavefunctions}

In RT-TDDFT, the typical problem is to evolve $N_{states}$ Kohn-Sham (KS) wavefunctions according to
\begin{equation}
	\frac{\partial }{\partial t} \psi_{j\mathbf{k}} (\mathbf{r},t) = -\mathrm{i} \hat{H}(\mathbf{r},t)\psi_{j\mathbf{k}} (\mathbf{r},t) \quad\quad j=1,\ldots,N_{states},
\end{equation}
where $\psi_{j\mathbf{k}} (\mathbf{r},t)$ is the wavefunction corresponding to the $j$-th KS state, and $\mathbf{k}$ is the electron wavevector. The time-dependent KS hamiltonian, $\hat{H}(\mathbf{r},t)$, can be written in the velocity gauge as:
\begin{equation}
	\hat{H}(\mathbf{r},t) = \frac{1}{2} \left[ -\mathrm{i}\nabla + \frac{\mathbf{A}(t)}{c} \right]^2 + V(\mathbf{r},t).
\end{equation}
Here, $c$ stands for the speed of light, $\mathbf{A}(t)$ for the vector potential of the applied electric field $\mathbf{E}(t)$, and $V(\mathbf{r},t)$ for the time-dependent KS potential.

%\subsubsection{Polarizability}\label{section:polarizability}
For molecules, a relevant physical quantity to monitor during the time evolution is the dipole moment $\bm\mu(t)$ \cite{Yabana_2006,Jornet-Somoza_2019,Andrade_2007}:
\begin{equation}\label{eq:dipole}
	\bm\mu(t) = -\int \mathbf{r} \ \! n(\mathbf{r},t)\mathrm{d}\mathbf{r} + \sum_\xi Z_\xi \mathbf{R}_\xi,
\end{equation}
where $\mathbf{R}_\xi$ is the position of nucleus $\xi$ with atomic number $Z_\xi$, and $n(\mathbf{r},t)$ is the electronic density. With periodic boundary conditions, $\bm\mu(t)$ as given in Eq. (\ref{eq:dipole}) is ill defined. Instead, changes in the dipole moment, $\delta\bm\mu(t)$, are well defined and can be easily calculated as
\begin{equation}
	\delta\bm{\mu}(t) = \bm{\mu}(t)-\bm{\mu}(0) = \Omega\int_{0}^{t} \mathbf{J}_{ind}(t') \mathrm{d}t',
\end{equation}
where $\Omega$ is the unit-cell volume, and $\mathbf{J}_{ind}(t)$, the induced current density, which is evaluated according to Ref. \cite{Pela_2021}.

By transforming $\mathbf{J}_{ind}(t)$ and $\mathbf{E}(t)$ to the frequency domain, the polarizability $\alpha$ can be obtained as
\begin{equation}\label{eq:polarizability}
	\alpha_{\lambda\lambda'}(\omega) = \frac{\mathrm{i}\Omega}{\omega}\frac{J_{\lambda}(\omega)}{E_{\lambda'}(\omega)},
\end{equation}
where $\lambda$ and $\lambda'$ represent the cartesian componenents $x,y,z$.

\subsection{LAPW+LO basis set}

In the LAPW+LO method, the unit cell is divided into two regions: non-overlapping spheres centered at each atom, named as muffin-tin (MT) spheres, and the interstitial (I) region. The KS wavefunctions are expanded as \cite{Gulans_2014}: 
\begin{equation}\label{eq:expansion_LAPW_lo}
    |\psi_{j\mathbf{k}} (t) \rangle = 
    \sum_{\mathbf{G}} C_{\mathbf{G}j,\mathbf{k}}(t) |\phi_{\mathbf{G}+\mathbf{k}} \rangle + 
    \sum_{\gamma} C_{\gamma j,\mathbf{k}}(t) |\phi_{\gamma} \rangle,
    \quad
    j=1,\ldots,N_{states},
\end{equation}
where $|\phi_{\mathbf{G}+\mathbf{k}} \rangle$ are augmented planewaves and  $|\phi_{\gamma} \rangle$ are LO's. The augmented planewaves (called LAPW's for simplicity) are represented as planewaves in I, which are smoothly augmented into the MT spheres in terms of atomic-like functions:
\begin{equation}
\phi_{\mathbf{G}+\mathbf{k}}(\mathbf{r}) = 
\left\{
\begin{array}{ll}
\displaystyle{\frac{1}{\sqrt{\Omega}}\mathrm{e}^{\mathrm{i}(\mathbf{G}+\mathbf{k})\cdot\mathbf{r}}},     &  \mathbf{r}\in \mathrm{I},\\
& \\
\displaystyle{\sum_{lm}\mathcal{A}_{\xi lm\mathbf{G},\mathbf{k}} u_{l \xi}^\mathcal{L}(r_{\xi})Y_{lm}(\hat{r}_{\xi})},     & \mathbf{r}\in \mathrm{MT}_\xi,
\end{array}
\right.
\end{equation}
while the LO's are non-zero only inside a specific MT sphere:
\begin{equation}
    \phi_{\gamma}(\mathbf{r}) = 
    u^\ell_{l\xi}(r_{\xi})Y_{l m}(\hat{r}_{\xi})\quad \mathbf{r}\in \mathrm{MT}_\xi.
\end{equation}
$\mathbf{G}$ is a reciprocal lattice vector, $Y_{lm}$ are spherical harmonics, $\mathcal{A}_{\xi lm\mathbf{G},\mathbf{k}}$ are the augmentation coefficients, $\mathbf{r}_{\xi}$ is an auxiliary variable: $\mathbf{r}_{\xi}= \mathbf{r}-\mathbf{R}_\xi$, and $u_{l \alpha}^\mathcal{L}(r_{\xi})$ and $u^\ell_{l\xi}(r_{\xi})$ are radial functions obtained from the solution of the radial Schr\"odinger equation (for details, see Ref. \cite{Gulans_2014}).

With the introduction of $\nu$ as a generic index for a basis element, Eq.(\ref{eq:expansion_LAPW_lo}) can be rewritten as:
\begin{equation} \label{eq:expansion_basis}
|\psi_{j\mathbf{k}} (t) \rangle = 
    \sum_{\nu=1}^{N_{basis}} C_{\nu j,\mathbf{k}}(t) |\phi_{\nu} \rangle,\qquad \quad j=1,\ldots,N_{states}
\end{equation}
where $N_{basis}=N_{LAPW}+N_{LO}$ is the total number of basis functions. The set of expansion coefficients for each \kpt\ can be condensed into a matrix $C_\mathbf{k}(t)$ of order $N_{basis} \times N_{states}$, which evolves in time as
\begin{equation}\label{eq:evolution_wf_coeff}
S_\mathbf{k} \dot{C}_\mathbf{k}(t) = -\mathrm{i}H_\mathbf{k}(t)C_\mathbf{k}(t).
\end{equation} 
$S_\mathbf{k}$ and $H_{\mathbf{k}}(t)$ are the overlap and Hamiltonian matrices.
% Equation (\ref{eq:evolution_wf_coeff}) can be evolved in time using a propagator $U(t)$ \cite{2006TDFT,2012FoTD,Castro_2004,GomesPueyo_2018}
% \begin{equation}
% C_\mathbf{k}(t+\Delta t) = U_\mathbf{k}(t) C_\mathbf{k}(t),
% \end{equation}
% that can be approximated in the simplest case as
% \begin{equation}\label{eq:propagator}
% U_\mathbf{k}(t) = \exp[-\mathrm{i}S^{-1}_\mathbf{k}H_{\mu\nu\mathbf{k}}(t)\Delta t].
% \end{equation}

\subsection{Auxiliary subspace}
The minimum number of basis functions required for a precise description of $|\psi_{j\mathbf{k}} (t) \rangle$ can be huge for supercells containing a vacuum layer, such as those required to study molecules, nanowires, or 2D systems. In such cases, a direct solution of Eq. (\ref{eq:evolution_wf_coeff}) may become cumbersome or even prohibitively expensive, given that it requires $\mathcal{O}(N^2_{basis}N_{states})$ FLOPs.

We now introduce the initial KS states (those at $t=0$) as an auxiliary basis, 
\begin{equation}\label{eq:auxiliary_basis}
\Large |\psi_{i\mathbf{k}}(0)\rangle, \qquad i=1,\ldots,N_{aux}.
\end{equation}
If $\Large |\psi_{j\mathbf{k}}(t)\rangle$ 
is well represented through a projection onto this auxiliary subspace, the approximation
\begin{equation}\label{eq:approximation_coefficients}
C_\mathbf{k}(t) \cong \mathbb{C}_\mathbf{k}(0)\theta_\mathbf{k}(t)
\end{equation}
holds. Here,
$\theta_\mathbf{k}(t)$ stands for a matrix of order $N_{aux} \times N_{states}$ with the projection coefficients $ \theta_{ij\mathbf{k}}(t) = \langle \psi_{i\mathbf{k}}(0) |\psi_{j\mathbf{k}}(t)\rangle$. $\mathbb{C}_\mathbf{k}(0)$ contains the expansion coefficients of the auxiliary functions defined in Eq. (\ref{eq:auxiliary_basis}) in terms of the LAPW+LO basis. The matrices $\mathbb{C}_\mathbf{k}(0)$ and $C_\mathbf{k}(t)$ coincide for a certain number of columns when $t=0$. However, we adopt different symbols to represent them, since $\mathbb{C}_\mathbf{k}(0)$ has $N_{aux}$ and $C_\mathbf{k}(t)$ has $N_{states}$  columns. 

The time evolution of $\dot{\theta}_\mathbf{k}(t)$ can be approximated as
\begin{equation}\label{eq:evolution_of_projection_coeff}
\dot{\theta}_\mathbf{k}(t)
\cong
-\mathrm{i}\mathbb{H}_\mathbf{k}(t)\theta_\mathbf{k}(t),
\end{equation} 
where the matrix $\mathbb{H}_\mathbf{k}(t)$, of order $N_{aux}\times N_{aux}$, is given by $H_\mathbf{k}(t)$ projected onto the auxiliary subspace
\begin{equation} \label{eq:definition_auxiliary_hamiltonian}
\mathbb{H}_\mathbf{k}(t) = \mathbb{C}_\mathbf{k}^\dagger(0)H_\mathbf{k}(t)\mathbb{C}_\mathbf{k}(0).
\end{equation}	
With the introduction of the auxiliary basis, $C_\mathbf{k}(t)$ can be obtained through Eq. (\ref{eq:approximation_coefficients}), while $\theta_\mathbf{k}(t)$ is propagated in time using Eq. (\ref{eq:evolution_of_projection_coeff}). Solving these equations requires $\mathcal{O}(N_{basis}N_{aux}N_{states})$ and $\mathcal{O}(N^2_{aux}N_{states})$ FLOPs, respectively. Thus, employing the auxiliary basis results in a more favorable procedure whenever $N_{aux} \ll N_{basis}$.

\subsection{Building $\mathbb{H}_\mathbf{k}(t)$ with submatrices}

Unlike Eq. (\ref{eq:definition_auxiliary_hamiltonian}) suggests, an explicit storage of $H_\mathbf{k}(t)$ is not required to obtain $\mathbb{H}_\mathbf{k}(t)$. The strategy to avoid it starts with expressing $\mathbb{H}_\mathbf{k}(t)$ as a sum:
\begin{equation}\label{eq:auxiliary_hamiltonian_with_parts}
\mathbb{H}_\mathbf{k}(t) = \mathbb{T}_\mathbf{k}+
\frac{\mathbf{A}(t)}{c}\cdot \mathbb{P}_\mathbf{k} + \frac{\mathbf{A}^2(t)}{2c^2}\mathbb{S}_\mathbf{k}+
\mathbb{V}_\mathbf{k}(t),
\end{equation}
where $\mathbb{T}_\mathbf{k}$, $\mathbb{P}_\mathbf{k}$, and $\mathbb{V}_\mathbf{k}(t)$ are matrices corresponding to the kinetic-energy, momentum, and KS potential operators, respectively, projected onto the auxiliary subspace. These projected matrices are related to their counterparts defined in the LAPW+LO basis, $T_\mathbf{k}$, $\mathbf{P}_\mathbf{k}$, and $V_\mathbf{k}(t)$, respectively, as:
\begin{eqnarray}
\mathbb{T}_\mathbf{k} &=& \mathbb{C}_\mathbf{k}^\dagger(0)T_\mathbf{k}\mathbb{C}_\mathbf{k}(0)
\nonumber
\\
\mathbb{P}_\mathbf{k} &=& \mathbb{C}_\mathbf{k}^\dagger(0)\mathbf{P}_\mathbf{k}\mathbb{C}_\mathbf{k}(0) 
\nonumber
\\
\mathbb{V}_\mathbf{k}(t) &=& \mathbb{C}_\mathbf{k}^\dagger(0)V_\mathbf{k}(t)\mathbb{C}_\mathbf{k}(0).
\end{eqnarray}
$\mathbb{V}_\mathbf{k}(t)$ is the only matrix in Eq. (\ref{eq:auxiliary_hamiltonian_with_parts}) whose elements must be recomputed at all time steps. Following \ref{sec:potential}, $\mathbb{V}_\mathbf{k}(t)$ can be decomposed as a sum of Interstitial and MT contributions, $\mathbb{V}^{\mathrm{I}}_\mathbf{k}(t)$ and $\mathbb{V}^{\mathrm{MT}}_\mathbf{k}(t)$, respectively:
\begin{equation}
\mathbb{V}_\mathbf{k}(t) = \mathbb{V}^{\mathrm{I}}_\mathbf{k}(t) + \mathbb{V}^{\mathrm{MT}}_\mathbf{k}(t) 
\end{equation}
with
\begin{equation}\label{eq:v_interstitial}
[\mathbb{V}^{\mathrm{I}}_\mathbf{k}(t)]_{ii'} = 
\sum_{\mathbf{G}\mathbf{G}'} 
C^*_{\mathbf{G}i\mathbf{k}}(0)
\tilde{V}_{\mathbf{G}-\mathbf{G}'}(t)
C_{\mathbf{G}'i'\mathbf{k}}(0), \quad i,i'=1,\ldots, N_{aux},
\end{equation}
\begin{eqnarray}
\mathbb{V}^{\mathrm{MT}}_\mathbf{k}(t)
&=&
\left[
\begin{array}{c}
   \mathbb{C}_{\mathcal{L},\mathbf{k}}(0)  \\
   \mathbb{C}_{\ell,\mathbf{k}}(0)
\end{array}
\right]^\dagger
\left[
\begin{array}{cc}
   \mathcal{A}_\mathbf{k}  & \mathbb{O} \\
   \mathbb{O}  & \mathbb{I}
\end{array}
\right]^\dagger
\left[
\begin{array}{cc}
   \mathcal{V}_{\mathcal{L}\mathcal{L}}(t)  & \mathcal{V}^T_{\ell\mathcal{L}}(t) \\
   \mathcal{V}_{\ell\mathcal{L}}(t)  & \mathcal{V}_{\ell\ell}(t)
\end{array}
\right]
\left[
\begin{array}{cc}
   \mathcal{A}_\mathbf{k}  & \mathbb{O} \\
   \mathbb{O}  & \mathbb{I}
\end{array}
\right]
\left[
\begin{array}{c}
   \mathbb{C}_{\mathcal{L},\mathbf{k}}(0)  \\
   \mathbb{C}_{\ell,\mathbf{k}}(0)
\end{array}
\right]
\nonumber
\\
&=& 
\mathbb{C}^\dagger_{\mathcal{L},\mathbf{k}}(0)
\mathcal{A}^\dagger_\mathbf{k}
\mathcal{V}_{\mathcal{L}\mathcal{L}}(t)
\mathcal{A}_\mathbf{k}
\mathbb{C}_{\mathcal{L},\mathbf{k}}(0) + 
\mathbb{C}^\dagger_{\mathcal{L},\mathbf{k}}(0)
\mathcal{A}^\dagger_\mathbf{k}
\mathcal{V}^T_{\ell\mathcal{L}}(t)
\mathbb{C}_{\ell,\mathbf{k}}(0) + \nonumber \\
&+&
\mathbb{C}^\dagger_{\ell,\mathbf{k}}(0)
\mathcal{V}_{\ell\mathcal{L}}(t)
\mathcal{A}_\mathbf{k}
\mathbb{C}_{\mathcal{L},\mathbf{k}}(0) 
+
\mathbb{C}^\dagger_{\ell,\mathbf{k}}(0)
\mathcal{V}_{\ell\ell}(t)\mathbb{C}_{\ell,\mathbf{k}}(0).\label{eq:v_MT}
\end{eqnarray}
The submatrices $\mathcal{V}_{\mathcal{L}\mathcal{L}}(t)$, $\mathcal{V}_{\ell\mathcal{L}}(t)$, $\mathcal{V}_{\ell\ell}(t)$, and $\tilde{V}_{\mathbf{G}-\mathbf{G}'}(t)$ are defined in \ref{sec:potential}. In Eq. (\ref{eq:v_MT}), the LAPW and LO elements of $\mathbb{C}_{\mathbf{k}}(0)$ are split into the submatrices $\mathbb{C}_{\mathcal{L},\mathbf{k}}(0)$, and $\mathbb{C}_{\ell,\mathbf{k}}(0)$, respectively. Equations (\ref{eq:auxiliary_hamiltonian_with_parts})-(\ref{eq:v_MT}) show that it is possible to evaluate $\mathbb{H}_\mathbf{k}(t)$ without explicit calculation of $H_\mathbf{k}(t)$.

Without using the auxiliary basis set, building and storing $H_\mathbf{k}(t)$ is necessary to evolve $C_\mathbf{k}(t)$ as in Eq. (\ref{eq:evolution_wf_coeff}). Setting up all its elements requires $\mathcal{O}(N_AN_{basis}^2)$ FLOPs, where $N_A$ is the number of augmentation coefficients for each APW function. In this process, one of the most costly parts is to carry out the product $\mathcal{A}^\dagger_\mathbf{k}\mathcal{V}_{\mathcal{L}\mathcal{L}}(t)\mathcal{A}_\mathbf{k}$, involving $\mathcal{O}(N_AN_{LAPW}(N_A+N_{LAPW}))$ FLOPs. On the other hand, using the auxiliary subspace permits to evaluate and store $\mathcal{A}_\mathbf{k} \mathbb{C}_{\mathcal{L},\mathbf{k}}(0)$ (\eg as a matrix $R$) with $\mathcal{O}(N_AN_{LAPW}N_{aux})$ FLOPs, and then perform $R^\dagger \mathcal{V}_{\mathcal{L}\mathcal{L}}(t)R$ with the cost of $\mathcal{O}(N_AN_{aux}(N_A+N_{aux}))$ FLOPs. Since $N_{LAPW}$ is typically much larger than $N_A$, building $\mathbb{H}_\mathbf{k}(t)$ instead of $H_\mathbf{k}(t)$ reduces the number of FLOPs by a factor of $\mathcal{O}(N_{LAPW}/N_{aux})$, which is very favorable when $N_{aux}\ll N_{LAPW}$.

%\subsubsection{Matrix-Matrix multiplication and Fast Fourier Transform}
The sum over $\mathbf{G'}$ in Eq. (\ref{eq:v_interstitial}) can be carried out with two approaches:
\begin{enumerate}
    \item Matrix-matrix (MM) multiplications: By converting $\tilde{V}_{\mathbf{G}-\mathbf{G}'}(t)$ into a matrix of order $N_{LAPW}\times N_{LAPW}$, and then performing a matrix-matrix multiplication with $\mathbb{C}_{\mathbf{k}}(0)$. This process costs $\mathcal{O}(N_{LAPW}^2N_{aux})$ FLOPs.
    \item Fast Fourier Transform (FFT): By recognizing the product as a convolution, a Fourier transformation of $\tilde{V}_{\mathbf{G}-\mathbf{G}'}(t)$ and $\mathbb{C}_{\mathbf{k}}(0)$ from reciprocal to real space can be employed, then a simple product of the two transformed functions can be carried out, which is then transformed back to reciprocal space. Taking advantage of FFT algorithms, this option requires $\mathcal{O}(N_{LAPW}N_{aux}\log(N_{LAPW}))$ FLOPs.
\end{enumerate}
Even though FFT presents better scaling than MM, according to our observations, there is a hidden prefactor that favors MM for small values of $N_{LAPW}$ and $N_{aux}$.

\section{Examples of applications}\label{sec:results}

We have implemented the formalism outlined in Section \ref{sec-theory} in \exciting, referred, from now on, as \newrt{}. In this Section, we validate it by using H$_2$ and CO as first benchmarks and use \exciting{} \fluorine\ as reference to compare the results. Then we present an example where a conventional calculation would be particularly expensive: It concerns a hybrid system consisting of a pyridine monolayer on MoS$_2$, which is excited by a laser pulse.

To evaluate the effectiveness of \newrt{}, we have performed timing measurements using a Dell workstation with an Intel Xeon W-2125 CPU and 64~GB of RAM. The Intel Fortran compiler, the Intel MPI implementation, and the MKL library as available in OneApi 2022.1 have been used to generate the executables. In all measurements, the CPU frequency was set to 1.2 GHz.

\subsection{H$_2$ dimer}

With only two electrons, H$_2$ is the simplest two-atomic molecule, representing the easiest test case for the new implementation. We adopt the experimental bond length of 1.401~bohr \cite{HCP92}, and place the molecule along the $x$ direction in a box of $21.401 \times 20 \times 20$~bohr$^3$. The dimensionless parameter \rgkmax{}, which controls the number of LAPWs, is set to 1.95 to keep the computational cost of the conventional implementation in \exciting\ \fluorine{} moderate; with sphere radii of $R_{MT}=0.65$~bohr, this results in $N_{basis}=3951$ basis functions. The dimension of the auxiliary subspace, $N_{aux}$, is given by $N_{empty}+1$, where $N_{empty}$ is the number of unoccupied states, an input parameter of \exciting{}.

\begin{figure}[htb]
	\centering
	\includegraphics[scale=1]{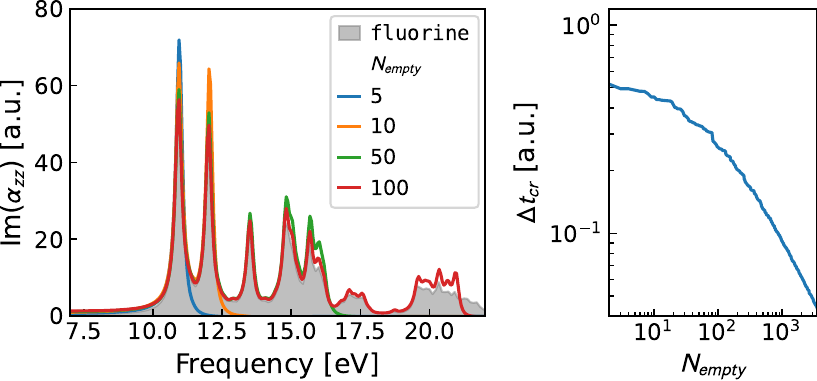}
	\caption{Imaginary part of the $zz$ component of the dynamical polarizability (left) for different values of $N_{empty}$. The shaded area shows the reference obtained with \exciting\ \fluorine. The right panel shows the critical time step to evolve the KS wavefunctions. Adopting $\Delta t\le \Delta t_{cr}$ ensures that the time evolution does not diverge.}
	\label{fig:H2-nempty}
\end{figure}

By applying an impulsive electric field along the $z$ direction, the dynamical polarizability $\alpha_{zz}$ is evaluated following \ref{sec-theory}.
Figure \ref{fig:H2-nempty} depicts on the left the imaginary part of the polarizability $\mathrm{Im}(\alpha_{zz})$ for different values of empty states. The reference spectrum, obtained with \fluorine{}, is depicted in shaded gray. The agreement improves by increasing $N_{empty}$, matching the reference up to higher and higher energies. For instance, with $N_{empty}=5$, 10, 50, and 100, good agreement is found up to approximately 11, 12, 16, and 21~eV, respectively.

Table \ref{tab:H2} shows the average time needed for one time step $\Delta t$  to evolve the KS wavefunctions as a function of $N_{empty}$ and the method used to evaluate Eq. (\ref{eq:v_interstitial}) (MM and FFT). For the selected values of $N_{empty}$, the times with the new implementation are shorter by factors of 4 to 139 compared to \fluorine. Furthermore, MM is faster than FFT for $N_{empty}\le 100$.

\begin{table}[htb]
    \centering
    \caption{Timing of one time step $\Delta t$ to evolve the KS system for the example of H$_2$ for different values of $N_{empty}$ and using the two different convolution methods (MM and FFT). The timings presented here have been scaled by 0.13~s. The critical time step $\Delta t_{cr}$ in a.u., obtained with Eq. (\ref{eq:critical_timestep}), is also shown, together with the overall speedup with respect to the conventional implementation.}
    \label{tab:H2}
    \begin{tabular}{c|r|SS|S|rr}
    \hline
    & & \multicolumn{2}{c|}{Timing} & & \multicolumn{2}{c}{Speedup} \\
     & $N_{empty}$& MM & FFT & $\Delta t_{cr}$ &  MM & FFT \\ \hline
    \multirow{4}{*}{\newrt{}} & 5 & 1.0 & 2.3& 0.49 & 56758 & 24678\\
    & 10 & 1.0 & 2.3 & 0.45 & 52125 & 22663\\
    & 50 & 1.5 & 16.9 & 0.33 & 25483 & 2261\\
    & 100 & 33.5 & 33.5 & 0.26 & 899 & 899 \\
%    \hline \hline
%        & & \multicolumn{2}{c}{Timing} & $\Delta t_{cr}$ & \multicolumn{2}{c}{Speedup} \\
        & & \multicolumn{2}{c|}{} &  & \multicolumn{2}{c}{} \\
    \hline
%    \fluorine &  & \multicolumn{2}{c}{139.0} & 0.0012 & \multicolumn{2}{c}{1} \\ \hline
    \fluorine &  & \multicolumn{2}{c|}{139.0} & 0.0012 & \multicolumn{2}{c}{} \\ \hline
    \end{tabular}
\end{table}

Figure \ref{fig:H2-nempty} depicts on the right an estimate of the critical time step $\Delta t_{cr}$ for evolving the KS wavefunctions. $\Delta t_{cr}$ has been calculated so that Eq. (\ref{eq:evolution_of_projection_coeff}) or Eq. (\ref{eq:evolution_wf_coeff}) for \exciting\ \fluorine{} does not diverge if the time step satisfies $\Delta t \le \Delta t_{cr}$:
\begin{equation}\label{eq:critical_timestep}
\Delta t_{cr} = \frac{f}{\epsilon_{max}-\epsilon_{min}}.
\end{equation}
$\epsilon_{max}$ ($\epsilon_{min}$) represents the maximum (minimum) eigenvalue of $\mathbb{H}_\mathbf{k}$. $f$ is a scaling factor related to how conservative the estimate should be (here $f=0.2$). For larger values of $N_{empty}$, $\Delta t_{cr}$ becomes smaller, because $\epsilon_{max}$ increases as more states are included. As can be seen in Fig. \ref{fig:H2-nempty}, for $N_{empty}$ being very large, $\Delta t_{cr}$ decreases as a power law of $N_{empty}$. The actual values of $\Delta t_{cr}$ are collected in Table \ref{tab:H2}. For the conventional implementation, the eigenvalues of $H_\mathbf{k}$ (not $\mathbb{H}_\mathbf{k}$) are used to evaluate $\Delta t_{cr}$. This yields a $\Delta t_{cr}$ that is more than 200 times smaller than all others.

With $\Delta t_{cr}$ and the timings to evolve the KS wavefunctions, the overall speedup of \newrt{} with respect to \fluorine{} can be determined. This is done by considering how long it takes to evolve the KS wavefunctions along a time window of 1 a.u. With the observed speedups listed in Table \ref{tab:H2}, \newrt{} can be faster than \fluorine{} by factors of $~$900-50,000.

\subsection{CO molecule}
CO is an experimentally relevant molecular probe to study the optical properties of more complex materials \cite{Chen_2003,Kustov_1991,Bordiga_2005}. Being a simple molecule, it has often been used to test new developments of methods that address excited states \cite{Gavnholt_2008,Zhang_2020,Hirose_2015,Sitt_2007}. This is also the aim here. In our calculations, the experimental bond length of 2.13 bohr \cite{HCP92} is used, and $20$~bohr of vacuum is adopted along each cartesian direction. The size of the auxiliary subspace, $N_{aux}$, is set to $N_{empty}+5$.

\subsubsection{Polarizability}
\begin{figure}[htb]
	\centering
	\includegraphics[scale=1]{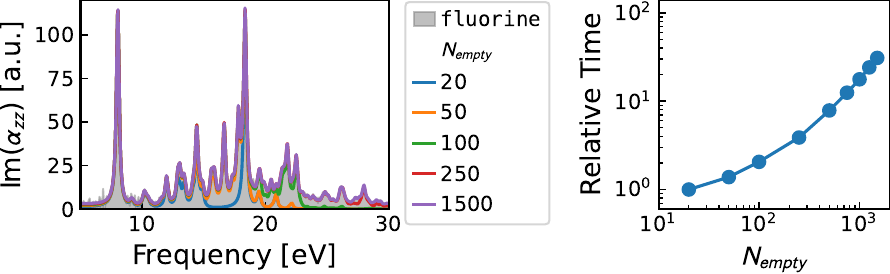}
	\caption{Left: impact of $N_{empty}$ on the polarizability of the CO molecule. The curves, obtained with the new algorithm for different values of the parameter $N_{empty}$, are compared to the standard approach (gray area). Right: Time relative to the case $N_{empty}=20$ (0.028~s) taken in a single iteration to evolve the KS wavefunctions.}
	\label{fig:CO-nempty}
\end{figure}

Initially, CO is subjected to a laser pulse with an impulsive electric field along the $z$ direction (perpendicular to the molecular axis). Figure \ref{fig:CO-nempty} displays on the left $\mathrm{Im}(\alpha_{zz})$ for different values of $N_{empty}$, with the reference calculation shown in shaded gray. As for H$_2$, increasing $N_{empty}$ improves the spectra, at the higher computational cost  (see right panel). It is apparent that already $N_{empty}=20$ is sufficient to converge the polarizability up to 15 eV. Increasing $N_{empty}$ to 50 converges the spectra up to approximately 20 eV. With a value of 100, the energy range of $0-23$~eV is well described. Finally, 250 empty states are sufficient to cover the entire energy range up to 30 eV. Increasing $N_{empty}$ to 1500 does not show any improvement in this energy window.

\subsubsection{Exposure to a laser}

\begin{figure}[htb]
	\centering
	\includegraphics[scale=1]{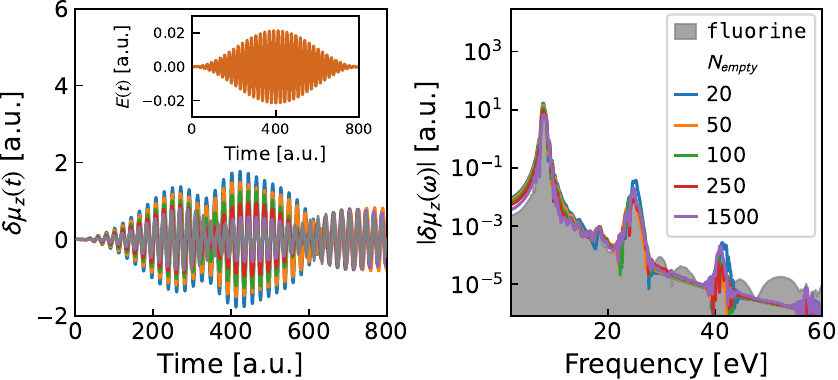}
	\caption{Change in the dipole moment of CO along the $z$ direction in time (left) and frequency (right) domain upon exposure to a laser (inset on the right).}
	\label{fig:CO-laser-jind}
\end{figure}

\begin{figure}[htb]
	\centering
	\includegraphics[scale=1]{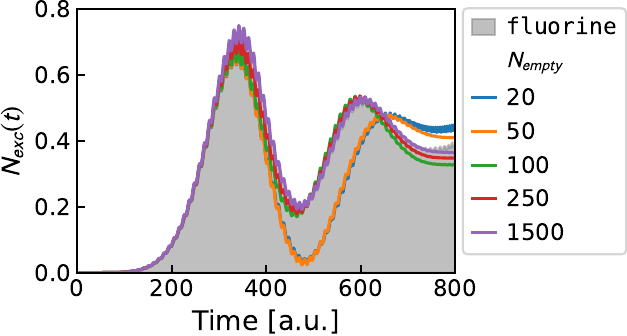}
	\caption{Number of electrons in CO excited by a laser pulse.}
	\label{fig:CO-laser-nexc}
\end{figure}

In a second step, CO is exposed to a laser whose field is described by the vector potential $\mathbf{A}(t)=A_0\cos(\omega t)\sin^2[\pi(t-t_0)/T_p] \hat{e}_z$, where $A_0=10$~a.u., $T_p=800$~a.u., $t_0=1$~a.u., and $\omega$ is 8~eV. The electric field $E(t)$ is illustrated in the inset of Fig. \ref{fig:CO-laser-jind} (left). It approximates a pulse with a Gaussian envelope, as is very common in experiments. The changes in the dipole moment, $\delta \mu_z$, are depicted in Fig. \ref{fig:CO-laser-jind}: on the left, the time evolution of $\delta \mu_z$; on the right, $\delta\mu_z$ transformed to the frequency domain.
As $N_{empty}$ is increased, the time-evolution $\delta \mu_z(t)$ approaches the result obtained with \fluorine{}, although a perfect match is hardly achieved, even when $N_{empty}=1500$. On the other hand, in the frequency domain (right panel), both results agree well, even for the smallest value of $N_{empty}=20$. The number of electrons excited due to the laser pulse is shown in Fig. \ref{fig:CO-laser-nexc}. Except for the smallest values of $N_{empty}$, \ie 20 and 50, all curves exhibit excellent agreement with the standard implementation in \fluorine.

\subsubsection{Increasing the basis-set size}
\begin{figure}[htb]
	\centering
	\includegraphics[scale=1]{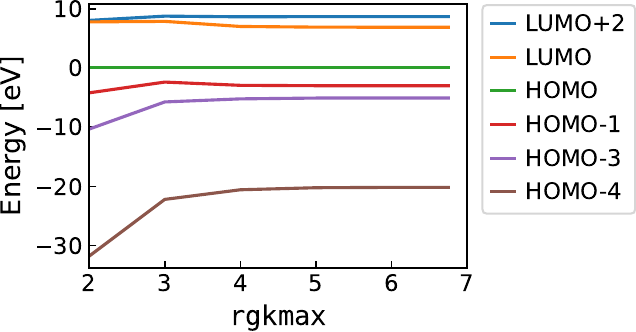}
	\caption{Convergence behavior of the KS energy levels with respect to the dimensionless parameter \rgkmax, which controls the number of basis functions. The HOMO level is set to zero. The lowest unoccupied molecular orbital (LUMO) and LUMO+1  as well as HOMO-1 and HOMO-2 are degenerate states.}
	\label{fig:CO-energy-levels}
\end{figure}

In \exciting{}, the dimensionless parameter \rgkmax{} defines a cutoff for LAPW's to be included in a calculation, thereby controlling the basis-set size. Figure \ref{fig:CO-energy-levels} shows how \rgkmax{} impacts the position of the KS levels in CO, where the energy of the highest occupied molecular orbital (HOMO) is used as a reference. An asymptotic convergence is observed upon increase of \rgkmax{}. Obviously, small values like \rgkmax{=2} or 3 are insufficient, while steady convergence can be seen for larger values, \ie \rgkmax{} of 4 (5) guarantees KS levels to be converged within 410 (65) meV with respect to results of the best calculations (\rgkmax=6.75).

Increasing \rgkmax{} also means increasing the computational effort. Table \ref{tab:CO-timing} shows how \rgkmax{} affects the number of basis functions ($N_{basis}$) and the corresponding effect on the time needed to evolve KS wavefunctions. 
\begin{table}[htb]
\centering
\caption{Influence of the basis-set size on the computational cost required to evolve KS wavefunctions. The parameter \rgkmax{} governs the number of basis functions $N_{basis}$ and thus heavily impacts the timing. The memory requirements for a calculation in the standard implementation with \rgkmax{=5} were more than 64~GB and thus beyond the computational resources of the used workstation. }\label{tab:CO-timing}
\begin{tabular}{crrrr}
\hline
& & \multicolumn{3}{c}{Timing [s]} \\
\rgkmax{} & $N_{basis}$ & \fluorine & MM & FFT \\
\hline
2.0 & 1723 & 1.8 & 0.11 & 0.11 \\
3.0 & 5591 & 45.5 & 0.90 & 3.66 \\
4.0 & 13195 & 476.3 & 4.86 & 3.88\\
5.0 & 25727 &  & 18.63 & 4.14 \\
\hline
\end{tabular}
\end{table}
For all calculations with \newrt{}, $N_{empty}$ was set to 250. With \fluorine{}, the calculation with \rgkmax{=5} required more memory than available on our workstation. For \rgkmax{=4}, a speedup of 98 (123) over \fluorine{} is observed with \newrt{} when MM (FFT) is employed. Interestingly, MM scales approximately with the square of $N_{basis}$, as we would expect if the convolution operation is the most time-consuming task. With FFT, there is no clear scaling behavior. For small \rgkmax\ (values of 2 and 3, \ie underconverged), MM shows better timings than FFT, whereas the opposite happens for larger \rgkmax\ values. This suggests that FFT will be the most appropriate option for production calculations.

\subsection{Hybrid interface pyridine@MoS$_2$}
To demonstrate the power of \newrt{}, we use it in an example of a typical real-world application. Inspired by Ref. \cite{Oliva_2022}, we consider here the interface composed of a monolayer of pyridine (C$_5$H$_5$N) molecules and 2D MoS$_2$, denoted as pyridine@MoS$_2$. Such hybrid materials receive increasing attention due to their potential applications in optoelectronic devices, such as photodetectors, photovoltaic absorbers, and nanoscale transistors \cite{Oliva_2022,Birmingham_2019,Klots_2014,Manzeli_2017,Kwon_2019,Lembke_2015}.

\begin{figure}[htb]
	\centering
	\includegraphics[scale=0.3]{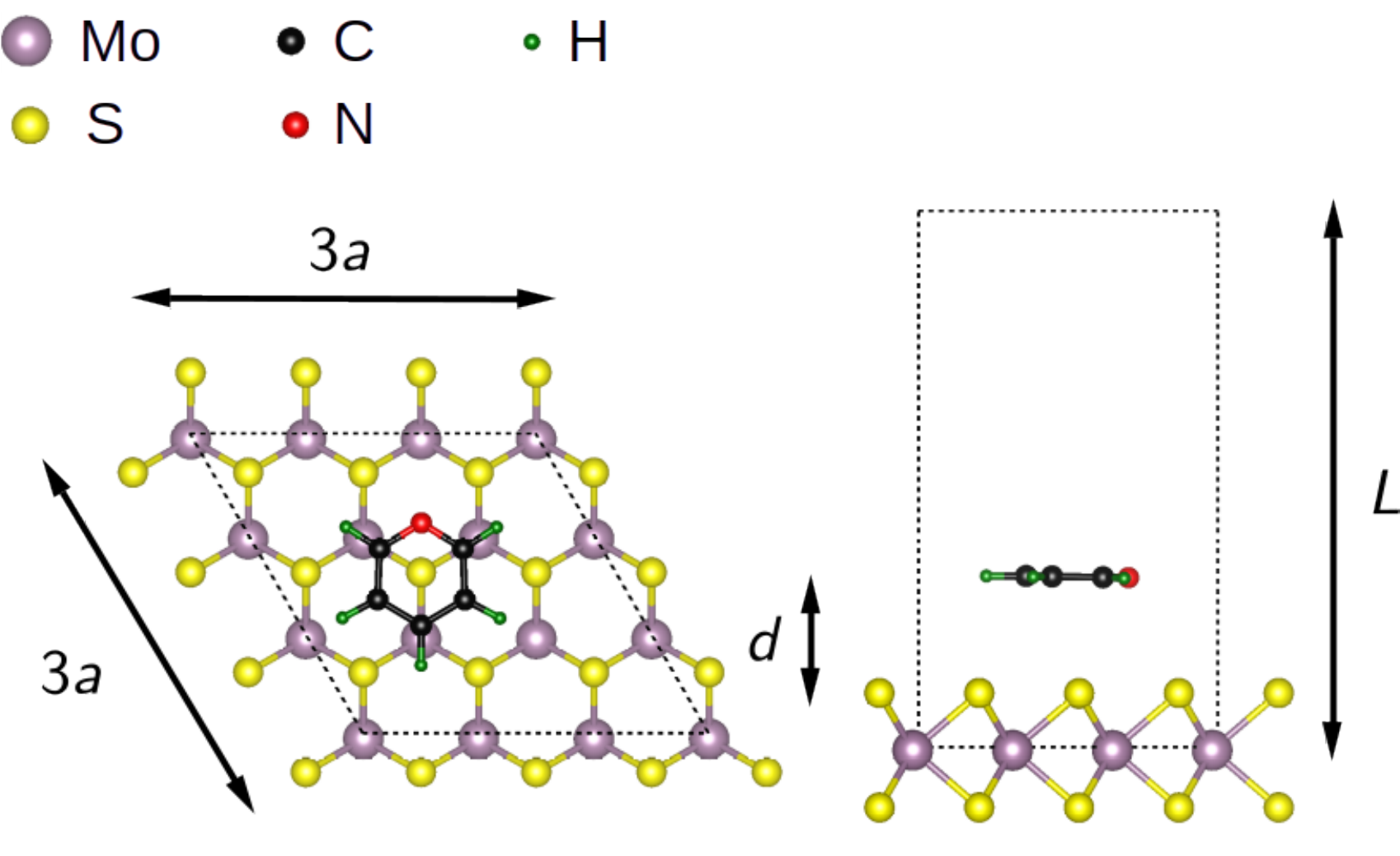}
	\caption{Unit cell used to study the interface of monolayer MoS$_2$ and the organic molecule pyridine (C$_5$H$_5$N). The length $L$ along the direction perpendicular to the monolayer plane is 14.71~\AA.}
	\label{fig:cell}
\end{figure}	

The structure is illustrated in Fig. \ref{fig:cell}; details are  reported in Ref. \cite{Oliva_2022}, and the input and output data \cite{Oliva_2022_nomad} are stored in the NOMAD data infrastructure \cite{Draxl_2019,Draxl_2018}. Pyridine@MoS$_2$ is realized as a supercell of lateral dimensions of $3a \times 3a$, where $a=3.16$~\AA{} is the experimental lattice parameter of MoS$_2$ \cite{Boeker_2001}. Pyridine is placed at a distance of $d=3.19$~\AA{} above the top sulfur atom. To isolate the neighboring replica along the $z$ direction, a supercell length of $L=14.71$~\AA{} is adopted. A $3\times 3 \times 1$ \kgrid{} is used, and the dimensionless parameter \rgkmax{} is set to 5. As the auxiliary basis of groundstate KS wavefunctions, we consider 132 occupied and 101 unoccupied states at each {\bf k}-point, covering the energy range from $-57$ to $11$ eV. Pyridine@MoS$_2$ is exposed to a laser pulse with a of duration 400~a.u, a frequency of 3.3~eV, which is modulated by a sin-squared function. This frequency is chosen based on the dielectric function of pyridine@MoS$_2$ \cite{Oliva_2022}, which exhibits a peak around 3.3~eV \cite{Oliva_2022} in the independent-particle approximation. The vector potential $\mathbf{A}$ of the laser pulse is parallel to the MoS$_2$ plane, with a maximum amplitude of 10~a.u. ($2.4\times10^{12}$~W/cm$^2$).  A time step of $\Delta t=0.02$~a.u. is used to evolve the KS wavefunctions. 

\begin{figure}[htb]
	\centering
	\includegraphics[scale=1]{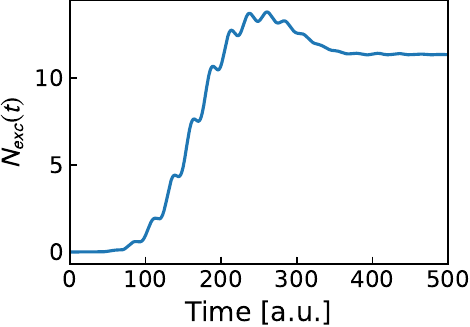}
	\caption{Number of excited electrons in one unit cell of pyridine@MoS$_2$.}
	\label{fig:MoS2-pyridine-nexc}
\end{figure}	

Figure \ref{fig:MoS2-pyridine-nexc} shows how the number of excited electrons $N_{exc}(t)$ increases over time until it reaches a maximum, and then finally decreases to a steady value of 11.3 for $t\ge 400$~a.u., when the laser pulse is not present any more. Considering that the supercell accommodates 9 unit cells of pristine MoS$_2$, this means 1.26 electrons in each of these 9 cells. 

\begin{figure}[htb]
	\centering
	\includegraphics[scale=1]{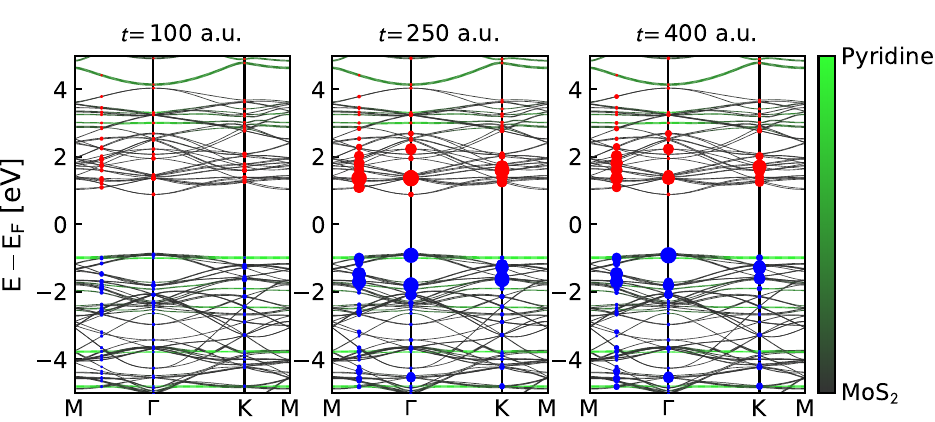}
	\caption{Bandstructure of pyridine@MoS$_2$, with the number of excited electrons (holes) in the conduction (valence) bands at times $t=$100 (left), 250 (middle), and 400~a.u. (right), respectively, depicted as red (blue) circles. The projection of the KS states onto MoS$_2$ and pyridine is represented according to the color map.}
	\label{fig:MoS2-pyridine-bands}
\end{figure}	
The distribution of the excited electrons in \textbf{k}-space is displayed in Fig. \ref{fig:MoS2-pyridine-bands} for times of 100, 250, and 400~a.u. The bands are projected onto MoS$_2$ and pyridine, and the color map indicates the projection: Green (black) indicates states of pyridine (MoS$_2$) character. The excitations are depicted as colored circles whose area is proportional to the number of electrons or holes in the corresponding state (in red and blue, respectively). Comparing the three times, the number of excitations is relatively small at $t=100$~a.u., it is the largest at $t=250$~a.u., and it decreases by a small amount at $t=400$~a.u. The majority of excitations are distributed over MoS$_2$, while only a small fraction of holes (electrons) can be observed at the pyridine HOMO (LUMO).

\section{Conclusions}
In this manuscript, we have presented a numerical approach for speeding up RT-TDDFT calculations in the all-electron full-potential code \exciting{}, targeted at systems with large unit cells. By employing an auxiliary subspace and submatrices, the number of FLOPs can be significantly reduced, while, simultaneously, allowing for larger simulation time steps. We have validated this procedure with the examples of H$_2$ and CO, showing that the accuracy of the calculations can be controlled by the dimension of the auxiliary subspace. Two algorithms, namely matrix-matrix multiplications and Fast Fourier Transforms have been introduced to project the auxiliary hamiltonian onto the auxiliary subspace. The first one has turned out more adequate for systems with small basis sets, whereas the second one is better suited for systems with large basis sets and is expected to be more suitable in production calculations. To prove the efficiency of the new implementation, we have applied it also to pyridine@MoS$_2$, a prototypical example of a real-world application, studying its excitations when exposed to a laser pulse. All input and output files underlying this publication are available at NOMAD  \cite{Draxl_2019,Draxl_2018} through the following link: \url{https://dx.doi.org/10.17172/NOMAD/2024.03.06-1} \cite{nomad-doi}. Our developments will be part of the next release of \exciting{} (sodium).

\section{Acknowledgements}
This work was supported by the German Research Foundation within project Nr. 182087777 - SFB 951 (CRC HIOS). Partial funding is appreciated from the European Union’s Horizon 2020 research and innovation program under the grant agreement Nº 951786 (NOMAD CoE). The authors gratefully acknowledge the computing time on the high-performance computer ``Lise'' at the NHR Center NHR@ZIB. This center is jointly supported by the Federal Ministry of Education and Research and the state governments participating in the NHR (\url{www.nhr-verein.de/unsere-partner}).

\appendix

\section{Expressions for the potential matrix}\label{sec:potential}

Initially, the KS potential is decomposed into contributions from the interstitial region and the muffin-tin spheres as
\begin{equation}
V(\mathbf{r},t)= V^{\mathrm{I}}(\mathbf{r},t)+
V^{\mathrm{MT}}(\mathbf{r},t), 
\end{equation}
\begin{equation}
V^{\mathrm{I}}(\mathbf{r},t) = 
\left\{
\begin{array}{ll}
V(\mathbf{r},t),     & \mathbf{r} \in \mathrm{I} \\
0,     & \mathbf{r} \in \mathrm{MT}_\xi
\end{array}
\right.
\end{equation}
\begin{equation}
V^{\mathrm{MT}}(\mathbf{r},t) = 
\left\{
\begin{array}{ll}
0,     & \mathbf{r} \in \mathrm{I} \\
V(\mathbf{r},t),     & \mathbf{r} \in \mathrm{MT}_\xi
\end{array}
\right.
\end{equation}
With this decomposition, the LAPW-LAPW, LO-LAPW, and LO-LO elements of the matrix $V_\mathbf{k}(t)$ are, respectively,
\begin{equation}
\langle \phi_{\mathbf{G}+\mathbf{k}} | V(\mathbf{r},t)| \phi_{\mathbf{G}'+\mathbf{k}} \rangle =  
\left[\mathcal{A}_{\mathbf{k}}^\dagger 
\mathcal{V}_{\mathcal{L}\mathcal{L}}(t)\mathcal{A}_{\mathbf{k}}\right]_{\mathbf{G},\mathbf{G}'} + \tilde{V}_{\mathbf{G}-\mathbf{G}'}(t),
\end{equation}
\begin{equation}
\langle \phi_{\gamma} | V(\mathbf{r},t) | \phi_{\mathbf{k}+\mathbf{G}'}\rangle =
\left[ \mathcal{V}_{\ell\mathcal{L}}(t)\mathcal{A}_{\mathbf{k}}\right]_{\gamma,\mathbf{G}'},
\end{equation}
\begin{equation}
\langle \phi_{\gamma} | V(\mathbf{r},t) | \phi_{\gamma'}\rangle = 
\left[ \mathcal{V}_{\ell\ell}(t)\right]_{\gamma,\gamma'},
\end{equation}
where $\tilde{V}_{\mathbf{G}-\mathbf{G}'}(t)$ is the Fourier transform of $V^{\mathrm{I}}(\mathbf{r},t)$
\begin{equation}
\tilde{V}_{\mathbf{G}-\mathbf{G}'}(t) = \frac{1}{\sqrt{\Omega}} \int \mathrm{d}\mathbf{r} V^{\mathrm{I}}(\mathbf{r},t) \mathrm{e}^{-\mathrm{i}(\mathbf{G}-\mathbf{G}')\cdot\mathbf{r}}.
\end{equation}
The matrices $\mathcal{V}_{\mathcal{L}\mathcal{L}}(t)$, $\mathcal{V}_{\ell\mathcal{L}}(t)$, and $\mathcal{V}_{\ell\ell}(t)$ contain integrals of the KS potential evaluated in the MT spheres. Taking $\nu$ as an index that stands for an element of a basis function inside an MT sphere ($\nu$ could come from an LAPW or a LO), the element $\nu\nu'$ of $\left[\mathcal{V}(t)\right]$ is non-zero only when $\nu$ and $\nu'$ are in the same MT $\xi$:
\begin{equation}
\left[\mathcal{V}(t)\right]_{\nu\nu'} 
=  
\sum_{LM} 
\mathcal{G}^{LM}_{l m , l' m'}  
\int_0^{R_{MT,\xi}} 
u^{\nu}_{l \xi}(r) 
v_{LM\xi}(r,t) 
u^{\nu'}_{ l' \xi}(r) r^2   
\mathrm{d}r,
\end{equation}
where $\mathcal{G}^{LM}_{l m , l' m'}$ are the Gaunt coefficients and $v_{LM\xi}(r_\xi,t)$ are determined from the expansion of $V^{\mathrm{MT}}(\mathbf{r},t)$
\begin{equation}
V^{\mathrm{MT}}(\mathbf{r},t)=\sum_{LM} v_{LM\xi}(r_\xi,t) y_{LM}(\hat{r}_\xi) \qquad \mathbf{r} \in MT_\xi,
\end{equation}
$y_{LM}$ being real spherical harmonics. 
%

%\section*{References}
%\bibliographystyle{unsrt}
%\bibliographystyle{iopart-num}
\bibliographystyle{elsarticle-num} 
\bibliography{references}

\end{document}